\documentclass{article}
\usepackage{spconf,amsmath,graphicx,hyperref,amsfonts}
\usepackage{multirow}
\usepackage{booktabs}   % for \toprule, \midrule, \bottomrule
\usepackage{amsmath}    % for math mode arrows (\uparrow, \downarrow)
\usepackage{amsmath,amsfonts,bm}

\def\eqref#1{equation~\ref{#1}}
\def\1{\bm{1}}

\def\vf{{\bm{f}}}

\def\vm{{\bm{m}}}

\def\vp{{\bm{p}}}

\def\vw{{\bm{w}}}
\def\vx{{\bm{x}}}

\def\vz{{\bm{z}}}

\DeclareMathAlphabet{\mathsfit}{\encodingdefault}{\sfdefault}{m}{sl}
\SetMathAlphabet{\mathsfit}{bold}{\encodingdefault}{\sfdefault}{bx}{n}

\title{MAGE: A Coarse-to-fine speech enhancer with Masked Generative Model}

\name{The Hieu Pham$^1$$^{\star}$, Tan Dat Nguyen$^2$$^{\star}$, Phuong Thanh Tran$^1$,  Joon Son Chung$^2$, Duc Dung Nguyen$^1$\thanks{$\star$ These authors contributed equally to this work.}}
  
  \address{$^1$AITech Lab, Ho Chi Minh City University of Technology, VNUHCM, Vietnam \\
      $^2$Korea Advanced Institute of Science and Technology, South Korea}
      
\begin{document}
%\ninept
%
\maketitle

\begin{abstract}
Speech enhancement remains challenging due to the trade-off between efficiency and perceptual quality. In this paper, we introduce \textbf{MAGE}, a \textbf{M}asked \textbf{A}udio \textbf{G}enerative \textbf{E}nhancer that advances generative speech enhancement through a compact and robust design. Unlike prior masked generative models with random masking, MAGE employs a scarcity-aware coarse-to-fine masking strategy that prioritizes frequent tokens in early steps and rare tokens in later refinements, improving efficiency and generalization. We also propose a lightweight corrector module that further stabilizes inference by detecting low-confidence predictions and re-masking them for refinement. Built on BigCodec and finetuned from Qwen2.5-0.5B, MAGE is reduced to 200M parameters through selective layer retention. Experiments on DNS Challenge and noisy LibriSpeech show that MAGE achieves state-of-the-art perceptual quality and significantly reduces word error rate for downstream recognition, outperforming larger baselines. Audio examples are available at \url{https://hieugiaosu.github.io/MAGE}.
\end{abstract}

\section{Introduction}
\label{sec:intro}

Speech enhancement (SE)~\cite{Richter2023SGMSE,Lemercier2023storm} is a fundamental task in speech processing that seeks to separate clean and intelligible signals from audio corrupted by diverse distortions.
These degradations arise from background noise, reverberation, microphone clipping, transmission artifacts, and resampling effects, and they can severely impair both human perception and machine processing. Achieving reliable SE is therefore crucial for a wide range of downstream applications, including personalized enhancement~\cite{parnamaa24_interspeech}, robust speech recognition~\cite{iwamoto22_interspeech}, and scalable data collection~\cite{Sabra2024SECP}.

Modern approaches fall broadly into two categories. \textit{Discriminative models} directly map noisy inputs to clean signals by optimizing losses aligned with intrusive metrics such as SI-SDR, PESQ, and STOI.
These models are computationally efficient, with DPRCN~\cite{le21dpcrnenhancment} achieving \textit{PESQ} $2.86$ using only $0.8$M parameters and TF-GridNet~\cite{ZhongQiuWang2023TF_gridnet} reaching \textit{PESQ} $3.78$ and \textit{STOI} $0.994$ with $9.8$M parameters. Nonetheless, despite progress in augmentation and adversarial training~\cite{wang19b_interspeech}, their performance often degrades when acoustic conditions differ from training data.
\textit{Generative models} instead seek to model the underlying speech distribution. Diffusion- and flow-based approaches such as SGMSE~\cite{Richter2023SGMSE} and StoRM~\cite{Lemercier2023storm} improve perceptual quality by reversing the corruption process, yet they remain computationally heavy and data-demanding, even with efficiency-oriented variants like SpeechFlow~\cite{liu2024speechflow}.
Discrete representations facilitate language-model-driven SE, as demonstrated by SELM~\cite{wang2024selm}, which leverages WavLM~\cite{Chen2022WavLM}, and MaskSR~\cite{li2024MaskSR} in combination with DAC~\cite{kumar2023dac}. However, these systems are still large and impractical for deployment. Recently, mask generative models (MGMs) such as MaskSR and AnyEnhance~\cite{zhang2025anyenhanceunifiedgenerativemodel} have shown state-of-the-art results, but they typically require hundreds of millions to billions of parameters and rely on random masking strategies that introduce inefficiency and redundancy. This gap motivates the need for generative SE models that deliver both perceptual quality and efficiency.

\begin{figure}[t]
    \centering
    \includegraphics[width=1\linewidth]{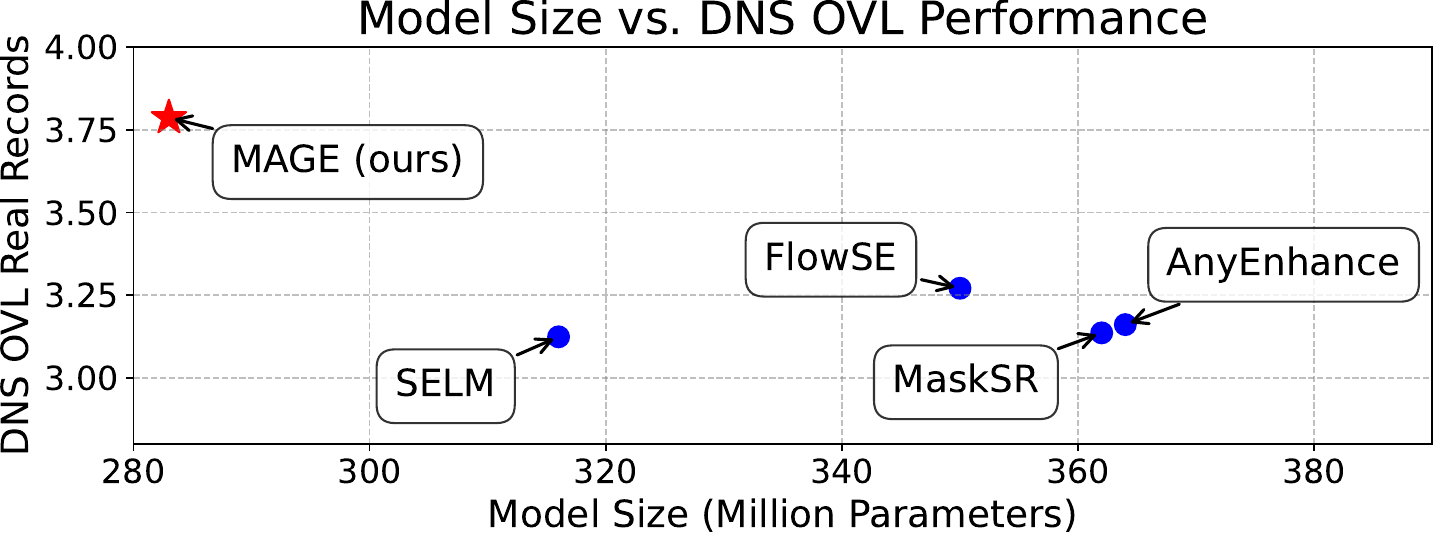}
    \caption{A comparison of DNS OVL scores on a real-world test set against model size for several different methods}
    \label{fig:compare_param_vs_dns}
\end{figure}

\begin{figure*}[t]
    \centering
    \includegraphics[width=\linewidth]{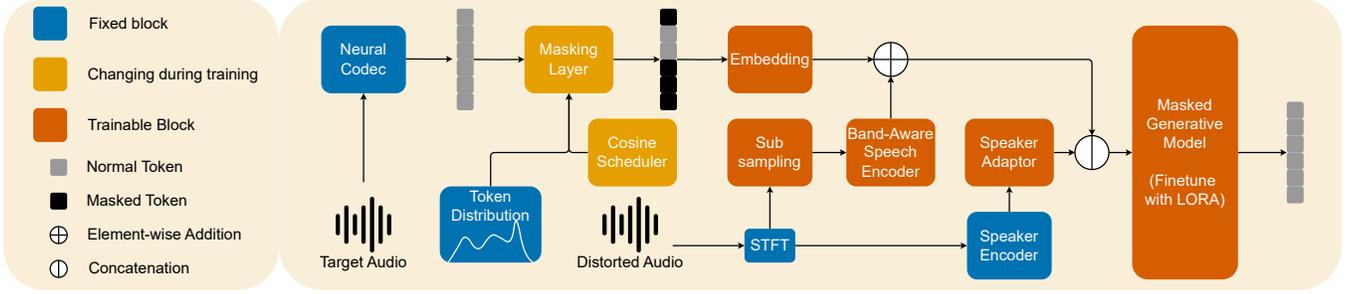}
    \caption{Training pipeline and model design of MAGE. Target audio is first converted into sequence of tokens using a Neural Encodec. These tokens are then masked according to their distribution to form a coarse-to-fine masking strategy, as described in Sec.~\ref{sec:coarse-to-fine}. Besides, speaker identity is extracted by a Band-Aware Encoder and a pretrained Speaker Encoder, enabling it to capture the acoustic characteristics. The model is optimized using cross-entropy loss applied only on the masked tokens.}
    \vspace{-1.77mm}
    \label{fig:MAGE}
\end{figure*}

In this work, we propose \textbf{MAGE}, a \textbf{M}asked \textbf{A}udio \textbf{G}enerative \textbf{E}nhancer that combines the perceptual advantages of generative modeling with the efficiency needed for practical deployment. Unlike prior mask generative models that rely on random masking and large parameter counts, MAGE introduces a scarcity-aware coarse-to-fine masking strategy and a lightweight corrector module, enabling more effective training and stable inference. Finetuned from Qwen2.5-0.5B and speech tokens from BigCodec, our proposed model is reduced to only 200M parameters while maintaining strong performance. Extensive experiments on DNS Challenge and noisy LibriSpeech show that MAGE achieves state-of-the-art perceptual quality and significantly improves word error rate (WER) for downstream automatic speech recognition, outperforming larger flow-based and mask-based baselines. These results highlight MAGE as a compact, generalizable, and robust framework for speech enhancement, advancing the development of resource-efficient generative models for real-world applications.

\section{MAGE}% Architecture}

\subsection{Neural Encodec and Encoders}
To extract discrete representations from target audio, we adopt BigCodec~\cite{ji2025wavtokenizer}, which provides stable tokenization and high-quality reconstruction using a single codebook with 80 tokens per second. To match this token rate, the distorted audio $\vw_{\text{distorted}}$ is converted into a complex spectrogram $\mathbf{S} \in \mathbb{C}^{F \times T}$ and downsampled with 2D convolutions. For conditioning, we leverage a TF-GridNet block~\cite{ZhongQiuWang2023TF_gridnet}, which efficiently models cross-band frequency interactions while remaining lightweight. The encoder output is projected into the embedding space $\vx_{\text{cond}} \in \mathbb{R}^D$ to condition the generative model. In addition, this complex spectrogram is used to extract speaker identity $\vx_e$ with a pretrained speaker encoder\footnote{\url{https://github.com/resemble-ai/Resemblyzer}}.

\subsection{Mask Generative Model (MGM)}
Given a token sequence $\vx=[x^1,\ldots,x^T]$ derived from $\vw_{\text{distorted}}$, the MGM learns to reconstruct masked tokens over $N$ denoising steps. At each step $i$, tokens are masked according to a binary vector $\vm_{(i)}$ sampled from a Bernoulli distribution:
\begin{equation} \label{eq:scheduler}
p(i)=\cos\!\left(\tfrac{\pi}{2}\tfrac{i}{N}\right),
\end{equation}
where $p(i)$ follows a cosine schedule~\cite{zhang2025anyenhanceunifiedgenerativemodel,li2024MaskSR}. We obtain the masked sequence $\tilde{\vx}_{(i)}$ by filling the mask token $M$ into the positions selected for masking, while keeping the remaining tokens unchanged. The masked sequence is then added element-wise with $\vx_{\text{cond}}$, and concatenated with the speaker embedding, which is projected from $\vx_e$ using a lightweight adaptor. The resulting sequence is used as the input to the masked generative model. The model parameters $\theta$ are optimized by predicting the masked tokens:
\begin{equation}
\mathcal{L}_{\text{mask}}=-\sum_{t=1}^T m^t_{(i)} \log P(x^t \mid \tilde{\vx}_{(i)}, \vx_{\text{cond}}, \vx_e; \theta).
\end{equation}
Intuitively, this objective forces the model to learn conditional distributions of tokens given both observed context and acoustic information.

\begin{table*}[t]
\centering
\small
\resizebox{1\textwidth}{!}{%
\begin{tabular}{lcccccccccccc}
\toprule
\multirow{2}{*}{\textbf{System}} & 
\multicolumn{4}{c}{\textbf{With Reverb}} & 
\multicolumn{4}{c}{\textbf{Without Reverb}} & 
\multicolumn{3}{c}{\textbf{Real Recordings}} \\
\cmidrule(lr){2-5} \cmidrule(lr){6-9} \cmidrule(lr){10-12}
& \textbf{SIG}$\uparrow$ & \textbf{BAK}$\uparrow$ & \textbf{OVL}$\uparrow$ &\textbf{SSIM}$\uparrow$ 
& \textbf{SIG}$\uparrow$ & \textbf{BAK}$\uparrow$ & \textbf{OVL}$\uparrow$ &SSIM$\uparrow$ 
& \textbf{SIG}$\uparrow$ & \textbf{BAK}$\uparrow$ & \textbf{OVL}$\uparrow$ \\
\midrule
BigCodec Resyn. GT &  4.473 & 4.471 & 4.190 & 0.857  & 4.473 & 4.471 & 4.190 & 0.857 & -- & -- & -- \\
\midrule
Noisy & 1.760 & 1.497 & 1.392 & --    & 3.392 & 2.618 & 2.483 & --    & 3.053 & 2.510 & 2.255 \\
Conv-TasNet~\cite{Luo19Convtasnet} & 2.415 & 2.710 & 2.010 & \textbf{0.939} & 3.092 & 3.341 & 3.001 & \textbf{0.945} & 3.102 & 2.975 & 2.410 \\
% Demucs~\cite{demucs2022} & 2.510 & 2.641 & 2.215 & 0.941 & 3.124 & 3.257 & 3.010 & 0.952 & 2.974 & 2.870 & 2.291 \\
% Inter-SubNet~\cite{intersubnet2023} & 2.651 & 2.581 & 2.362 & 0.933 & 3.458 & 3.821 & 3.099 & 0.967 & 3.258 & 3.574 & 2.806 \\
% CDiffuSE~\cite{Lu2022CDifuSE} & 2.541 & 2.300 & 2.190 & 0.907 & 3.294 & 3.641 & 3.047 & 0.914 & 3.201 & 3.104 & 2.781 \\
SGMSE~\cite{Richter2023SGMSE} & 2.730 & 2.741 & 2.430 & 0.899 & 3.501 & 3.710 & 3.137 & 0.934 & 3.297 & 2.894 & 2.793 \\
StoRM~\cite{Lemercier2023storm} & 2.947 & 3.141 & 2.516 & \underline{0.934} & 3.514 & 3.941 & 3.205 & \underline{0.943} & 3.410 & 3.379 & 2.940 \\
% SELM (K=300)~\cite{Wang2024SELM} & 3.160 & 3.577 & 2.695 & 0.901 & 3.508 & 4.096 & 3.258 & 0.917 & 3.591 & 3.435 & 3.124 \\
ANYENHANCE~\cite{zhang2025anyenhanceunifiedgenerativemodel} & 3.500 & 4.040 & 3.204 & --    & 3.640 & 4.179 & 3.418 & --    & 3.488 & 3.977 & 3.161 \\
MaskSR-M~\cite{li2024MaskSR} & 3.531 & 4.065 & 3.253 & 0.827 & 3.586 & 4.116 & 3.339 & 0.929 & 3.430 & 4.025 & 3.136 \\
% MaskSR2-L~\cite{liu2024masksr2} & -- & -- & -- & -- & 3.638 & 4.153 & 3.400 & -- & 3.511 & 4.078 & 3.233 \\
FlowSE~\cite{flowse2025} & 3.614 & 4.110 & 3.340 & 0.809 & 3.690 & 4.200 & 3.451 & 0.940 & 3.643 & 4.100 & 3.271 \\
\midrule
\textbf{MAGE} & 3.530 & \textbf{4.149} & 3.107 & 0.724   & 4.407 & \underline{4.515} & 4.151 & 0.817    & 3.830 & \textbf{4.302} & 3.500 \\
+ Corrector & 3.525 & \underline{4.146} & 3.081 & 0.724 & 4.441 & \textbf{4.557} & 4.201 & 0.800 & 4.098 & \underline{4.309} & \underline{3.744} \\
+ CTF & \textbf{3.876} & 3.901 & \textbf{3.653} & 0.799 & \underline{4.559} & 4.408 & \textbf{4.235} & 0.819 & \textbf{4.206} & 4.145 & \textbf{3.787} \\
+ CTF \& Corrector & \underline{3.864} & 3.961 & \underline{3.372} & 0.789 & \textbf{4.580} & 4.338 & \underline{4.223} & 0.821 & \underline{4.191} & 3.924 & 3.666 \\
\bottomrule
\end{tabular}
}
\caption{Performance comparison on DNS Challenge test set. DNSMOS scores (SIG/BAK/OVL) and speaker cosine similarity (SSIM) are reported. Higher values ($\uparrow$) indicate better performance. BigCodec Resyn. GT denotes the ground-truth signal re-encoded and decoded to illustrate the upper bound of the method.}
% \vspace{-1.77mm}
\label{tab:dns_performance}
\end{table*}

% joonson@kaist.ac.kr

\subsection{Coarse-to-Fine (CTF) Strategy and Corrector}
\label{sec:coarse-to-fine}
Uniform masking in Eq.~\ref{eq:scheduler} treats all tokens equally, ignoring the fact that token frequencies are highly non-uniform. This creates a bias: frequent tokens dominate training, while rare tokens are predicted less reliably. In other words, the model learns 'popular' predictions more often than 'rare' predictions, which leads to suboptimal generalization.

To mitigate this imbalance, we introduce a \textbf{coarse-to-fine (CTF) masking strategy}. We compute a frequency vector $\vf=[f^1,\ldots,f^T]$ for each input $\vx$, where $f^i$ is the document frequency~\cite{tf-idf} of token $x^i$ in the training corpus. We then calculate an IDF-like score:
\begin{equation}
\vz = \log\!\left(\frac{N+1}{\vf+1}\right),
\end{equation}
where $N$ is the total number of samples of dataset. Higher $\vz$ indicates rarer tokens. The base masking probability for each token is
\begin{equation} \label{eq:p_base}
\vp_{\text{base}} = \sigma\!\left(\frac{\vz - \bar{\vz}}{\mathrm{std}(\vz)}\right),
\end{equation}
where $\sigma(\cdot)$ is the sigmoid function, and $\bar{\vz}$ and $\mathrm{std}(\vz)$ denote the mean and standard deviation of $\vz$. 
To preserve the advantages of the cosine schedule while making it token-dependent, we define the final CTF probability:
\begin{equation}
\vp_{\text{CTF}} = \min\!\left(\frac{E_{\text{cos}}}{E_{\text{base}}}\cdot\vp_{\text{base}},1\right),
\end{equation}
where $E_{\text{cos}}=T \sin\!\left(\tfrac{\pi i}{2N}\right)$ is the expected number of masked tokens under the cosine schedule, and $E_{\text{base}}=\sum_{t=1}^T p^t_{\text{base}}$ under the base distribution. This adjustment ensures that frequent tokens are predicted earlier, while rare tokens are emphasized in later steps, creating a natural curriculum.

\textbf{Corrector.} Finally, to further improve robustness, we add a correction stage. Unlike standard MGM which generates tokens once and commits to them, MAGE allows re-masking and regeneration of previously decoded tokens. This prevents error accumulation and enables iterative refinement. Instead of random re-masking as in MaskSR~\cite{li2024MaskSR}, we train a lightweight 4-layer BLSTM corrector that identifies low-confidence tokens and re-masks them. During training, up to 30\% of ground-truth tokens are randomly corrupted, teaching the corrector to detect inconsistencies. At inference, it selectively re-masks problematic tokens and passes them back to the generative model for correction. This design improves perceptual quality by enabling the model to self-revise.

\begin{table}[t]
\centering
\small
\resizebox{.48\textwidth}{!}{%
\begin{tabular}{lcccc}
\toprule
\multirow{2}{*}{\textbf{System}} & 
\multicolumn{4}{c}{\textbf{LibriSpeech}} \\
\cmidrule(lr){2-5}
& \textbf{SIG}$\uparrow$ & \textbf{BAK}$\uparrow$ & \textbf{OVL}$\uparrow$ & \textbf{WER}$\downarrow$ \\
\midrule
SGMSE \cite{Richter2023SGMSE} & 4.254 & 4.109 & 3.813 & 28.52\\
StoRM \cite{Lemercier2023storm} & 4.030 & 4.241 & 3.986 & 27.34 \\
FlowSE \cite{flowse2025} & 3.539 & 2.923 & 2.634 & 35.53 \\
\midrule
MAGE+CTF & 4.449 & \textbf{4.301} & 4.076 & 25.27 \\
MAGE+CTF+Corrector & \textbf{4.517} & \textbf{4.301} & \textbf{4.141} & \textbf{23.45} \\
\bottomrule
\end{tabular}
}
\caption{Performance on the noisy LibriSpeech test set. Reported metrics: DNSMOS (SIG/BAK/OVL, $\uparrow$) and WER ($\downarrow$).}
\label{tab:wer}
% \vspace{-1.77mm}
\end{table}

\section{Experimental Setup}
We finetune the masked language model from Qwen2.5-0.5B using LoRA~\cite{hu2022lora}. To reduce computational cost, only half of the original layers are retained, resulting in a compact model with 200M parameters. The MAGE speech encoder applies Short-time Fourier Transform with $n\_fft=256$, $window=256$, $hop\_size=100$, followed by two TF-GridNet blocks with embedding size of 48, BLSTM hidden size of 192, and 4 heads attention. The language model is a reduced Qwen2.5-0.5B, where only the odd-numbered layers are kept (layer 1 is the first), and attention is configured in a non-autoregressive mode. LoRA is applied to \texttt{q\_proj}, \texttt{v\_proj}, \texttt{o\_proj}, \texttt{up\_proj}, and \texttt{down\_proj} with $r=16$, $\text{lora\_alpha}=32$, and dropout 0.1. Training is performed with AdamW (learning rate and weight decay $1\times10^{-4}$), batch size 8, on a single RTX 4090 GPU.

The training corpus is constructed by augmenting clean speech from LibriSpeech~\cite{Panayotov2015librispeech} and the DNS Challenge~\cite{strake20_interspeech} with noise from WHAM!~\cite{wichern19wham} and DNS Challenge, and reverberation from OpenSLR28. The final dataset contains 512k four-second utterances at 16 kHz, with a composition of 50\% noise-only, 30\% noise+reverb, and 20\% noise+reverb combined with resampling and spectrogram augmentation.

We evaluate MAGE against both discriminative and generative baselines using standard multiple metrics: SIG (signal distortion), BAK (background intrusiveness), and OVL (overall quality) from ITU-T P.835 \cite{dnsmos}, as well as Speaker Similarity (SSIM), computed as the cosine similarity between speaker embeddings from Wespeaker \cite{wang2023wespeaker}. Results are reported under three conditions: with reverberation, without reverberation, and on real recordings. Higher values indicate better performance for all metrics. We also use the released ASR model\footnote{\url{https://huggingface.co/nvidia/stt_en_conformer_transducer_xlarge}} to compute WER, reflecting intelligibility gains.

\section{Results and Analysis}

\begin{table}[t]
\centering
\small
\resizebox{.48\textwidth}{!}{%
\begin{tabular}{lccc}
\toprule
\multirow{2}{*}{\textbf{System}} & 
\multicolumn{3}{c}{\textbf{Without Reverb}} \\
\cmidrule(lr){2-4}
& \textbf{SIG}$\uparrow$ & \textbf{BAK}$\uparrow$ & \textbf{OVL}$\uparrow$ \\
\midrule
SSL Model (HuBERT) \cite{hsu2021hubert} & 4.362 & \textbf{4.518} & \textbf{4.220} \\
Transformer (8 layers) & 3.414 & 4.002 &  3.276 \\
Transformer (6 layers) & 3.322 & 3.842 & 3.155 \\
\midrule
Band Aware & \textbf{4.407} & 4.515 & 4.151 \\
\bottomrule
\end{tabular}
}
\caption{Ablation study on different Speech Encoder}
\label{tab:ablation_speech_encoder}
% \vspace{-1.77mm}
\end{table}

Table~\ref{tab:dns_performance} compares MAGE with both discriminative and generative baselines on the DNS Challenge test set. Conventional discriminative models such as Conv-TasNet achieve strong SSIM scores but lag behind in perceptual quality (SIG/BAK/OVL). Generative methods including SGMSE, StoRM, and FlowSE offer improvements, yet remain limited in overall robustness. Mask-based approaches such as MaskSR-M and ANYENHANCE further advance BAK and OVL, but require considerably larger model sizes. In contrast, \textbf{MAGE} achieves competitive or superior performance with only 200M parameters. Notably, incorporating the coarse-to-fine (CTF) masking strategy yields substantial gains, especially in OVL (up to 4.235 without reverb and 3.787 on real recordings). The corrector module further stabilizes inference, and the combined CTF+Corrector configuration achieves the best SIG score of 4.580 without reverb. For reference, the \textit{BigCodec Resyn. GT} row reports scores obtained by reconstructing ground-truth clean audio directly through the codec, representing the upper bound of codec fidelity, especially speaker similarity.

Table~\ref{tab:wer} reports results on the noisy LibriSpeech benchmark, highlighting downstream ASR performance. While prior generative models such as SGMSE and StoRM reduce word error rate (WER) relative to the noisy baseline, \textbf{MAGE with CTF and Corrector} achieves a WER of 23.45\%, a relative improvement of over 5\% absolute compared to SGMSE. These findings confirm that the enhanced audio produced by MAGE not only improves perceptual quality but also provides tangible benefits for recognition tasks.

\begin{figure}[t]
    \centering
    \includegraphics[width=1\linewidth]{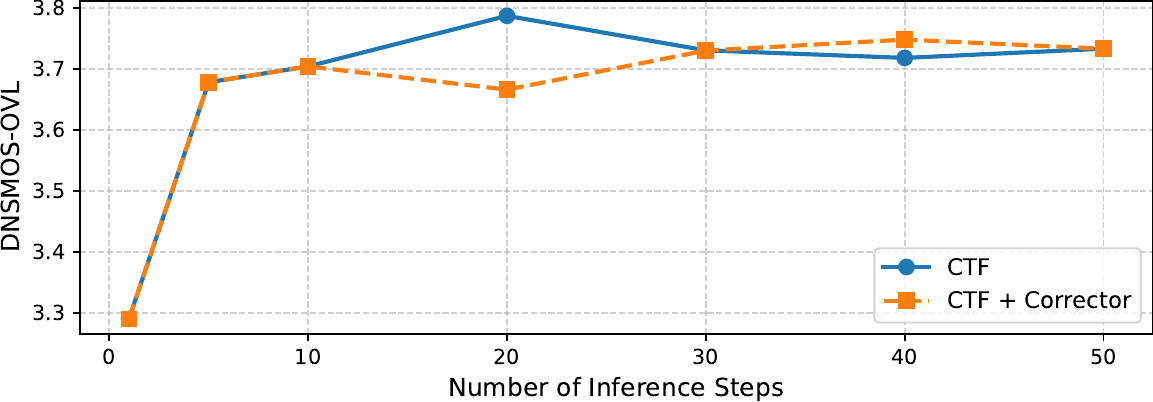}
    \caption{Ablation study on the number of inference steps for CTF and CTF + Corrector. 
The overall performance is measured using DNSMOS-OVL on Real Recording DNS dataset}
    \label{fig:inferstep}
    % \vspace{-1.77mm}
\end{figure}

Table~\ref{tab:ablation_speech_encoder} presents an ablation on the choice of speech encoder. SSL models, such as HuBERT, deliver strong performance but are computationally intensive. Lightweight transformer encoders reduce complexity but exhibit degraded quality. Howerver, the band-aware TF-GridNet encoder achieves a balance, reaching SIG 4.407, BAK 4.515, and OVL 4.151, comparable to HuBERT while being more efficient. This demonstrates that explicitly modeling cross-band spectral dependencies is more parameter-efficient than adopting SSL encoders. Fig.~\ref{fig:inferstep} presents an ablation on the number of inference steps for CTF and CTF+Corrector, evaluated on real recordings with DNSMOS-OVL. Performance improves rapidly up to 10 steps and stabilizes beyond 20 steps, with both variants maintaining strong quality across the range. CTF alone reaches a peak at 20 steps, while CTF+Corrector provides more stable performance at higher step counts, indicating that corrective refinement helps mitigate error accumulation without requiring excessive iterations.

Overall, these results show that MAGE effectively narrows the gap between perceptual quality and efficiency. The combination of CTF masking and corrective refinement delivers consistent improvements across datasets and metrics, establishing MAGE as a compact yet robust generative framework for speech enhancement. While MAGE achieves strong results, some limitations remain. Its dependence on simulated distortions may reduce generalization to real-world conditions, and the evaluation focuses mainly on DNSMOS instead of metrics such as PESQ or STOI.

\section{Conclusion}

In this work, we introduced MAGE, a lightweight difficulty-aware mask generative model for speech enhancement, finetuned from Qwen2.5-0.5B and coupled with BigCodec. Using a coarse-to-fine masking strategy with an auxiliary corrector, MAGE achieves strong perceptual quality and robustness with only 200M parameters. Experiments on DNS Challenge and LibriSpeech show that it outperforms discriminative and generative baselines in noisy and reverberant conditions. Future work includes extending to multilingual and streaming scenarios, joint training with ASR/TTS, and scaling to larger codecs and multimodal inputs, aiming to make MAGE a practical, generalizable solution for real-world deployment.

\newpage
% \begingroup
% \scriptsize
\bibliographystyle{IEEEbib}
% \fontsize{8.6pt}{8.6pt}\selectfont
\bibliography{shortstring,strings,refs}
% \endgroup

\end{document}